\documentclass[12pt]{article}
\usepackage{cite,epsfig,amssymb,amsmath,amstext,  graphicx,color,subfigure,xcolor,mathrsfs,tikz}
\evensidemargin \oddsidemargin

\usetikzlibrary{decorations.pathmorphing}
\usepackage{latexsym}
\usepackage{hyperref}
\hypersetup{colorlinks,%
  citecolor=blue,%
  linkcolor=red,%
  pdftex}
\usepackage[numbers,sort&compress,square]{natbib}
\usepackage{varioref}
\usepackage{slashed}
\usepackage{pdfsync}
\begin{document}

~
\vspace{5mm}

\begin{center}
{{\Large \bf Toward Quantization of Inhomogeneous Field Theory }}
\\[17mm]
O-Kab Kwon$^1$, ~~Jeongwon Ho$^2$, ~~Sang-A Park$^3$, ~~Sang-Heon Yi$^2$  \\[2mm]
{\it $^1$Department of Physics,~Institute of Basic Science, Sungkyunkwan University, Suwon 16419, Korea} \\
{\it $^2$Center for Quantum Spacetime, Sogang University, Seoul 04107, Republic of Korea} \\
{\it $^3$Department of Physics, Yonsei University, Seoul 03722, Korea} 
\\[2mm]
{\it okab@skku.edu, ~freejwho@gmail.com, ~psang314@gmail.com, ~shyi@sogang.ac.kr}
\end{center}
\vspace{15mm}

\begin{abstract}
\noindent
We explore the quantization of a $(1+1)$-dimensional inhomogeneous scalar field theory in which Poincar\'{e} symmetry is explicitly broken. We show the `classical equivalence' between a scalar field theory on curved spacetime background and its corresponding inhomogeneous scalar field theory.
This implies that a hidden connection may exist among some inhomogeneous field theories, which corresponds to general covariance in field theory on curved spacetime.
Based on the classical equivalence, we propose how to quantize a specific field theory with broken Poincar\'{e} symmetry inspired by standard field theoretic approaches, canonical and algebraic methods, on curved spacetime.
Consequently, we show that the Unruh effect can be realized in inhomogeneous field theory and propose that it may be tested by a condensed matter experiment.
We suggest that an algebraic approach is appropriate for the quantization of a generic inhomogeneous field theory.
\end{abstract}


\vskip 1.5in

\hrulefill

All authors contributed equally.

\thispagestyle{empty}
 
\newpage

\section{Introduction}

There are growing interests in physical systems described by particles or fields with non-constant mass, related to the condensed matter physics, neutron physics, cosmology and so on. For instance, in the study of electronic and transport properties of some materials, the Schr\"{o}dinger equation with effectively position-dependent mass is an important and useful tool~\cite{Bastard:1988,Roos:1983,Filho:2011}.  In nuclear physics and neutron star physics, this equation has also interesting applications~\cite{Ring:1980,Chamel:2005nd}.  In the perspective of field theory, the so-called waterfall field in a hybrid model of inflation~\cite{Linde:1993cn} can be regarded effectively as an incarnation of a time-dependent mass.  Recently, there have been some renewed interests in the supersymmetric extension of the position-dependent mass and couplings~\cite{Bak:2003jk,DHoker:2006qeo,Kim:2008dj,Kim:2009wv,Kim:2018qle,Kim:2019kns,Arav:2020obl,Kim:2020jrs,Kwon:2021flc}.  In these examples,  some quantum nature has been explored in the non-relativistic quantum mechanics setup and some classical aspects  in  field theories have  been analyzed.  Nevertheless,  the formalism and conceptual understanding of non-constant mass in field theories are  a bit deterred  partly because of the lack of sufficient symmetries or conservation properties. Furthermore, analytic solutions to equations of motion are  rare.   Accordingly, the quantum aspects of  field theory with  non-constant mass raises various issues and  awaits further explorations. 

In this paper, we consider a position-dependent mass and couplings in the context of bosonic field theories. We will call this kind of field theory as {\it inhomogeneous field theory} (IFT). Since the position dependence of mass and couplings breaks the Poincar\'{e} symmetry explicitly, the usual Wigner representation of the symmetry on fields would be  unavailable. In other words, scalar, vector or tensor field distinction or spin representation of fields are obscured in IFT, unlike in the  Poincar\'{e}   symmetric case. Despite this, we may simply take the standpoint that our IFT is derived from the usual field theory with constant mass and couplings  by elevating those to be position-dependent. Therefore, it would be legitimate to adopt the usual terminology and to designate a scalar field in IFT, for instance. This is an obvious  abuse of terminology, but wouldn't lead to any hurdles or glitches. 

On the other hand, a frame dependence might be a real trouble in quantizing IFT. One of the related hurdles in the quantization is the absence of Poincar\'{e} invariant vacuum, which will require  a careful treatment. This situation reminds us {\it field theory on curved spacetime} (FTCS), in which case one encounters similar issues like the absence of the preferred vacuum. In the case of FTCS, there has been a large development to overcome difficulties with preserving  general covariance, which is  eventually evolved to a subject known as the algebraic quantum field theory~\cite{Haag:1992hx,Wald:1995yp,Yngvason:2004uh,Halvorson:2006wj,Hollands:2009bke,Benini:2013fia,Hollands:2014eia,Khavkine:2014mta,Fredenhagen:2014lda}. See also~\cite{Witten:2018zxz,Witten:2021jzq,Dedushenko:2022zwd}. Based on the similar difficulties in IFT to those in FTCS, one may guess that it is very tempting to adopt the algebraic formulation for IFT. In particular, the van Hove model, which may be thought to be a kind of IFT, has been a good example for such a formulation~\cite{Fewster:2019ixc}. See also \cite{Wrochna:2011nx, Schlemmer:2015sna}.
 However, in general, it doesn't seem to be straightforward to materialize this expectation.

In this work, we would like to make this anticipation into a concrete working example in the case of a $(1+1)$-dimensional scalar IFT. Toward quantization of IFT, it would be an interesting direction inspired by an algebraic quantization process in FTCS. One of our main points  is to propose a way to quantize the IFT using a method that respects the classical equivalence between a FTCS and its corresponding IFT in $(1+1)$ dimensions. We also propose a ``generalized stress tensor'' in {\it inhomogeneous quantum field theory} (IQFT), which is designed to be conserved. Based on this, we show the existence of the Unruh-like effect in IQFT. Though one may regard our specific model  to have  a bit limited scope at this stage, our slogan is that an algebraic method is appropriate approach to a generic IQFT in any dimensions.

The paper is organized as follows. In section 2, we show the `classical equivalence' between a $(1+1)$-dimensional FTCS and its corresponding  $(1+1)$-dimensional IFT. As an example, we focus on a free scalar field.  The implication of general covariance in conjunction with the equivalence is explored. In some examples, the limiting behaviors of a position-dependent mass are inspected and their geometric interpretation is given. In section 3, based on the classical equivalence, we present our main proposal for the quantization of our model in IFT. Following the canonical quantization in curved spacetime, we quantize IFT by the canonical quantization. In section 4, the Hadamard method is summarized which is  a very useful algebraic construction superseding the canonical quantization. By applying this method to IFT, we suggest some interesting quantum aspects including the Unruh-like effect in IFT. In section 5, we summarize our results and provide some future directions. In Appendix \ref{AppA} and \ref{AppB}, we collect some formulae useful for the main text.

\section{Classical Relation between FTCS and IFT }

In conventional relativistic quantum field theories with constant mass and coupling parameters in Minkowski spacetime, the Poincar\'{e} symmetry is  essential in the canonical quantization. On the contrary,  the Poincar\'{e} symmetry is explicitly broken in IFT whose mass and coupling parameters depend on spatial coordinates. Therefore, it isn't guaranteed to  apply the canonical quantization method to IFT. In this paper, we would like to propose a  quantization method for  IFT, based on  a classical relation between    FTCS and IFT  in $(1+1)$-dimensional spacetime. Before presenting  our proposal for the quantization,  we  show  the classical relation  between FTCS and IFT in this section.
And we also discuss interesting implication of general covariance in FTCS in the context of IFT. 
Then, we investigate limiting behaviors of a position-dependent mass in terms of background metric in FTCS  through our classical relation.

\subsection{$(1+1)$-dimensional IFT and FTCS}

As a simple example of IFTs, let us consider a $(1+1)$-dimensional scalar field theory with a position dependent mass $m(x)$,  couplings $g_{n}(x)$'s, and a  source  $J(x)$, whose action is given by
\begin{equation} \label{IFT0}
S_{{\rm IFT}} =\int d^2 x \mathcal{L}_{{\rm IFT}} = \int d^2 x\left( -\frac{1}{2}\partial _{\mu}\phi \partial^{\mu}\phi - \frac{1}{2}m^{2}(x)\phi^{2} - \sum_{n=3}g_{n}(x)\phi^{n} +J(x)\phi\right)\,,
\end{equation}%
where  the Poincar\'{e} symmetry is partially broken. Using the remaining symmetry, its supersymmetric extension with $J(x)=0$ was done in \cite{Kwon:2021flc}. Attempting to quantize this theory,    we  encounter   various difficulties due to the  broken Poincar\'{e} symmetry.    Recalling that any parameters in field theory may be promoted to the background values of  certain fields in an enlarged field theory or string theory\footnote{This is an old folklore realized in various cases. For instance, axion field is a field elevation of the original constant $\theta$ angle parameter. The usefulness of space-dependent parameters in supersymmetric field theory is emphasized in~\cite{Seiberg:1993vc}.}, we may regard $m(x)$,  $g_{n}(x)$'s, and $J(x)$  as  vacuum expectation values of those fields.  For instance, we might embed the above action into a higher dimensional theory with the Poincar\'{e} symmetry. In this way, we may perform a quantization of IFT from the  Poincar\'{e} invariant theory. However, it doesn't seem to be  straightforward  to realize this embedding.  

Instead, we  explore another way for quantization of the above IFT. For this purpose,  we consider a $(1+1)$-dimensional scalar FTCS, whose action is given by
\begin{equation} \label{GraLag0}
S_{{\rm FTCS}} =\int d^2 x \sqrt{-g} \mathcal{L}_{{\rm FTCS}} = \int d^{2}x \sqrt{-g}\bigg[ -\frac{1}{2}\nabla _{\mu}\phi \nabla^{\mu}\phi - \frac{1}{2}m_0^{2} \phi^{2} - \sum_{\ell=1}f_{\ell}({\cal R})\phi^{\ell} \bigg]\,,
\end{equation}
where $m_0$ is a constant, ${\cal R}$ denotes the scalar curvature of the background metric, and $f_{\ell}$'s are  some functions of ${\cal R}$.
The action \eqref{GraLag0} enjoys general covariance. In the following, we  take the above metric, $g_{\mu\nu}$, as a non-dynamical one. Now, let us take a $(1+1)$-dimensional metric in the form of 
\begin{equation} \label{Ourmet0}
ds^{2} = e^{2\omega(x)}(-dt^{2} + dx^{2})\,,
\end{equation}
which is a generic conformal form of the metric in $(1+1)$ dimensions. The absence of the time coordinate in $\omega$ is matched to time independence of coupling parameters in IFT. See the action \eqref{IFT0}.  Some properties of this metric are given in Appendix~\ref{AppA}. In particular, one may note that $\sqrt{-g} = e^{2 \omega (x)}$ and ${\cal R} = -2 \omega'' e^{-2\omega}$,  where  ${}'$ denotes the derivative with respect to $x$.

Inserting the above metric in the FTCS action~\eqref{GraLag0}, one can see that the action is converted to the IFT action in \eqref{IFT0} with the parameter matching as follows:
\begin{align} \label{matching}
&m^{2}(x)  =  \sqrt{-g} \Big(m^{2}_{0} +2 f_{2}(\mathcal{R})\Big)\,,   
\nonumber\\
&
g_{n}(x) =  \sqrt{-g}   f_{n} (\mathcal{R})\,,  \nonumber\\
&J(x)  = -\sqrt{-g} f_{1}(\mathcal{R})\,.
\end{align}
Under this conversion, a  background metric is considered {\it  to be given}. Choosing $f_\ell(\mathcal{R})$'s, we obtain position-dependent parameters in IFT. Since $f_\ell(\mathcal{R})$'s can be chosen, in principle, as {\it arbitrary} functions of $\mathcal{R}$, a wide variety of IFTs can be considered in this way. 
For instance, even in the case that $x$-ranges of non-constant $m^2(x),\, g_n(x)$ and $J(x)$ do not overlap, we can find $f_\ell(\mathcal{R})$'s which are converted to those Lagrangian parameters.    

Under the conversion rule~\eqref{matching} for a {\it given} metric, the equations of motion of scalar fields are completely identical to each other in FTCS and IFT actions.\!\footnote{As described below, we use the term of  classical equivalence   within a class of IFT  such that the conversion is valid and the classical equations of motion of FTCS and IFT are identical.} Thus, we can say that the same physics is described by two differently-looking languages, FTCS and IFT, {\it i.e.}, a kind of dual description.
   We will call such a relation between two theories as the `classical equivalence' and denote:
\begin{equation}
        \boxed{\text{IFT $\iff$ FTCS}}\,.
\end{equation}
It is crucial in this equivalence that the kinetic term in \eqref{GraLag0} in $(1+1)$ dimensions is always independent of the metric for the conformal form in \eqref{Ourmet0}. The basic concepts and methods in the above can be extended to other field theories including fermionic ones in $(1+1)$ dimensions. However, note that in general the classical equivalence is not straightforwardly extended to higher dimensional theories. 
As alluded above, the `classical equivalence' does not mean that the theory spaces of FTCS and IFT are equivalent.  
In this paper, this term is used only in the  restricted sense that the conversion rules are performed and the corresponding IFT is determined. Though it may be regarded as nothing novel, it is shown to have notable and useful aspects, which are the main points in this paper. See also \cite{Ho:2022omx}.

To illustrate our approach concretely, we focus on the following simplest quadratic action on curved spacetime, which is  given by\footnote{This is not the conformally coupled case in $(1+1)$ dimensions.}
\begin{equation} \label{GraLag2}
S_{{\rm FTCS}}  =  \int d^{2}x \sqrt{-g} \Big[-\frac{1}{2}\nabla _{\mu}\phi \nabla^{\mu}\phi -\frac{1}{2}m^{2}_{0}\phi^{2} - \frac{\xi}{2} \mathcal{R} \phi^{2}\Big]\,,
\end{equation}
where $\xi$ is  a constant. The stress tensor for this action is given by
\begin{align} \label{}
T_{\mu\nu}
&= \nabla_{\mu}\phi\nabla_{\nu}\phi -\frac{1}{2}g_{\mu\nu}\Big[(\nabla\phi)^2 + m_{0}^2\, \phi^2\Big]  +  \xi  \Big(- \nabla_{\mu}\nabla_{\nu} + g_{\mu\nu}\nabla^{2} \Big)\phi^{2}\,.
\end{align}
Just like in the general case, one can see that the above FTCS action is converted to the IFT action on the Minkowski spacetime in the form of 
\begin{equation} \label{IFT2}
S_{{\rm IFT}} = \int d^{2}x \Big[ -\frac{1}{2}\partial _{\mu}\phi \partial^{\mu}\phi -\frac{1}{2}m^{2}(x)\,  \phi^{2}\Big]\,, \qquad 
\end{equation}
where the squared mass function is given by
\begin{equation} \label{Matrel}
m^2(x;m_0,\xi) = \sqrt{-g} (m_0^2 + \xi \mathcal{R})=e^{2\omega(x)}m_0^2-2\omega''\xi \,.
\end{equation}
When we consider the inverse conversion from IFT to FTCS, the choice of the function $f_{2}({\cal R})$ in~\eqref{matching} is not unique while keeping the equations of motion unchanged for a given  mass function $m(x)$. This means that we have freedom in the conversion process and these should be taken into account in the equivalence. Though a minimal choice is to take $\xi =0$, we consider a two-parameter conversion to set the scalar IFT action~\eqref{IFT2} to the form of \eqref{GraLag2}. This choice in \eqref{Matrel} contrasts to taking $m(x)$ as the vacuum expectation value of a certain scalar field and reveals some interesting  features in our equivalence, especially in the expression of ``stress tensor'' in IFT, $T_{\mu\nu}^{\rm IFT}$.

From the stress tensor in the scalar FTCS, we read the ``stress tensor'' in the scalar IFT as
\begin{align} \label{Tclassical}
T_{\mu\nu}^{\rm IFT} & = \partial_{\mu}\phi\partial_{\nu}\phi -\frac{1}{2}\eta_{\mu\nu}\Big[(\partial\phi)^{2} + m^{2}_{0}e^{2\omega}\phi^{2}\Big] +\xi\Big[-\partial_{\mu} \partial_{\nu}  +  \Gamma^{\alpha}_{\mu\nu,\,  {\rm IFT}} \partial_{\alpha}  + \eta_{\mu\nu} \partial^{2} \Big]\phi^{2}\,,
\end{align}
where $\Gamma^{\alpha}_{\mu\nu,\,  {\rm IFT}}$ denotes 
\begin{equation} \label{} 
\Gamma^{\alpha}_{\mu\nu,\,  {\rm IFT}} \equiv \omega'(x) \Big[2 \delta^{(x}_{\mu}\delta^{t)}_{\nu}  \delta^{\alpha}_{t}+ (\delta^{x}_{\mu} \delta^{x}_{\nu}  +  \delta^{t}_{\mu}\delta^{t}_{\nu})\delta^{x}_{\alpha}\Big] \,.
\end{equation}
Here, $\omega'(x)$ should read from the chosen conversion rule in~\eqref{Matrel}. We would like to emphasize that our ``stress tensor'' satisfies the ``conservation law'' derived from the counter part in FTCS, which takes a very unusual form from the viewpoint of IFT, {\it i.e.},
\begin{align}
        \nabla^{\mu}_{{\rm IFT}}T_{\mu\nu}^{\rm IFT} \equiv \eta^{\alpha\beta} \big[ \partial_{\alpha}T_{\beta\nu}^{\rm IFT} - \Gamma^{\rho}_{\alpha\beta,\, {\rm IFT}}~ T_{\rho\nu}^{\rm IFT}      - \Gamma^{\rho}_{\alpha\nu,\, {\rm IFT}}~ T_{\beta\rho}^{\rm IFT} \big] =0\,, 
\end{align}
where the indices are raised and lowered by $\eta_{\mu\nu}$ in IFT.  Indeed,   one can check explicitly in our case that
\begin{align}    \label{}
\nabla^{\rm IFT}_{\mu} T^{\mu}_{~\,t, {\rm IFT} } &=\partial_{\mu} T^{\mu}_{~\,t, {\rm IFT} } =0\,,  \nonumber \\
\nabla^{\rm IFT}_{\mu} T^{\mu}_{~\,x, {\rm IFT} } &=  \partial_{\mu} T^{\mu}_{~\,x, {\rm IFT} }  +\omega'(x) \Big[m^{2}_{0}e^{2\omega} -\xi\partial^2 \Big]\phi^{2}  =0\,.
\end{align}
The first equation corresponds to the energy conservation which comes from the $t$-translation symmetry in our scalar IFT. The second equation obviously reveals the absence of the $x$-translation symmetry in IFT, which implies that the conserved stress tensor cannot be introduced in the usual sense~\cite{Kwon:2021flc}. In other words, there is no conserved quantity coming from the second equation. Nevertheless we can interpret the above quantity as the conserved ``stress tensor'' originated from the scalar FTCS. 

To see legitimacy of our construction of the ``stress tensor'' and to verify the energy conservation, one may return to the canonical Hamiltonian. 
In our specific position dependent IFT in~\eqref{IFT2}, the canonical Hamiltonian (density) can be introduced as
\begin{equation} \label{}
{\cal H}_{\rm can} = \frac{1}{2}\Big[\dot{\phi}^{2} + (\phi')^{2}  + m^{2}(x) \phi^{2}\Big]\,,\qquad \dot{}\equiv\frac{\partial}{\partial t}\,,
\end{equation}
which corresponds to the time translation generator. To see the matching of this to the ``stress tensor'', let us check that 
  the $tt$-component of our ``stress tensor'' is the same with ${\cal H}_{\rm can}$ up to the total derivative term,
\begin{equation} \label{}
T^{tt}_{\rm IFT} 
= {\cal H}_{\rm can} - \Big[ \xi\Big ((\phi^{2})' -\omega'\phi^{2}\Big)\Big]' \,,
\end{equation}
where one should recall that the indices are raised by $\eta_{\mu\nu}$ in \eqref{Tclassical}. 
If the fall-off condition of the scalar field $\phi$ is taken appropriately, the energy defined by the Hamiltonian would be the same as that obtained from the $tt$-component of our ``stress tensor''.

As discussed previously, the classical equivalence provides us an interesting way to explore the quantization of the IFT with a position dependent mass. Before going ahead, we present some notable aspects of our model and its correspondence.

\subsection{Implications  of general covariance}

As we have already discussed,  general covariance is manifest in the quadratic FTCS action \eqref{GraLag2}. Though the equivalence of \eqref{GraLag2} and \eqref{IFT2} for the metric \eqref{Ourmet0} is shown, the meaning of general covariance is obscured in the IFT action point of view since the measures of the action integral may be changed by general coordinate transformations.   In this subsection, we try to figure out the implication of the general covariance of FTCS in the context of IFT in association with the classical equivalence.

As we showed in the previous subsection, the FTCS action \eqref{GraLag2} is converted to the IFT action \eqref{IFT2} as
\begin{align}\label{IFT=g1}
 \int d^2 x\, \sqrt{-g} \mathcal{L}_{{\rm FTCS}} (g_{\mu\nu}, m_0, \xi)= \int d^2 x\, \mathcal{L}_{{\rm IFT }} (m(x)),
\end{align}
where $m^2(x) = \sqrt{-g} (m_0^2 + \xi \mathcal{R})$ and the metric is given in \eqref{Ourmet0}. We would like to ask how the general covariance in the left-hand side is encoded in the right-hand side of \eqref{IFT=g1}.
Performing a general coordinate transformation $(t,x) \,\, \longrightarrow \,\, (T(x^\mu), X(x^\mu))$ for the FTCS action in the left-hand side of \eqref{IFT=g1}, we obtain
\begin{align}\label{GC}
 \int d^2 x\, \sqrt{-g} \mathcal{L}_{{\rm FTCS}} (g_{\mu\nu}, m_0, \xi)= \int d^2 X \sqrt{-\bar{g}}\, \mathcal{L}_{{\rm FTCS}} (\bar{g}_{\mu\nu}, m_0, \xi)\,,
\end{align}
where $\bar{g}_{\mu\nu} (X)\equiv \frac{\partial x^\rho}{ \partial X^\mu} \frac{\partial x^\sigma}{ \partial X^\nu} g_{\rho\sigma}(x)$. Under this transformation, the integration region could be changed. 
Now we convert the transformed FTCS action to the IFT action:
\begin{align}\label{IFT=g2}
\int d^2 X \sqrt{-\bar{g}}\, \mathcal{L}_{{\rm FTCS}} (\bar{g}_{\mu\nu}, m_0, \xi)= \int d^2 X \mathcal{L}_{{\rm IFT }} (\bar{m}(\mathbf{X})),
\end{align}
where the mass function in the right-hand side is given by
\begin{align}\label{mpX}
\bar{m}^2 (\mathbf{X})  =  \sqrt{-\bar{g}} (m_0^2 + \xi \bar{\mathcal{R}}).
\end{align} 
Compared to the position-dependent mass function $m(x)$, the mass function $\bar{m}(\mathbf{X})$ in the scalar IFT converted from the transformed scalar FTCS can depend on both coordinates $\mathbf{X}=(T,X)$. That is, two seemingly different IFT's with $m(x)$ and $\bar{m}(\mathbf{X})$ are connected each other. This result is quite surprising from the view point of IFT, while it is natural within our equivalence. Based on this observation, we would like to interpret this relation between two IFT's as the existence of a hidden connection in IFT.

\begin{figure}
        \begin{center}
        \includegraphics{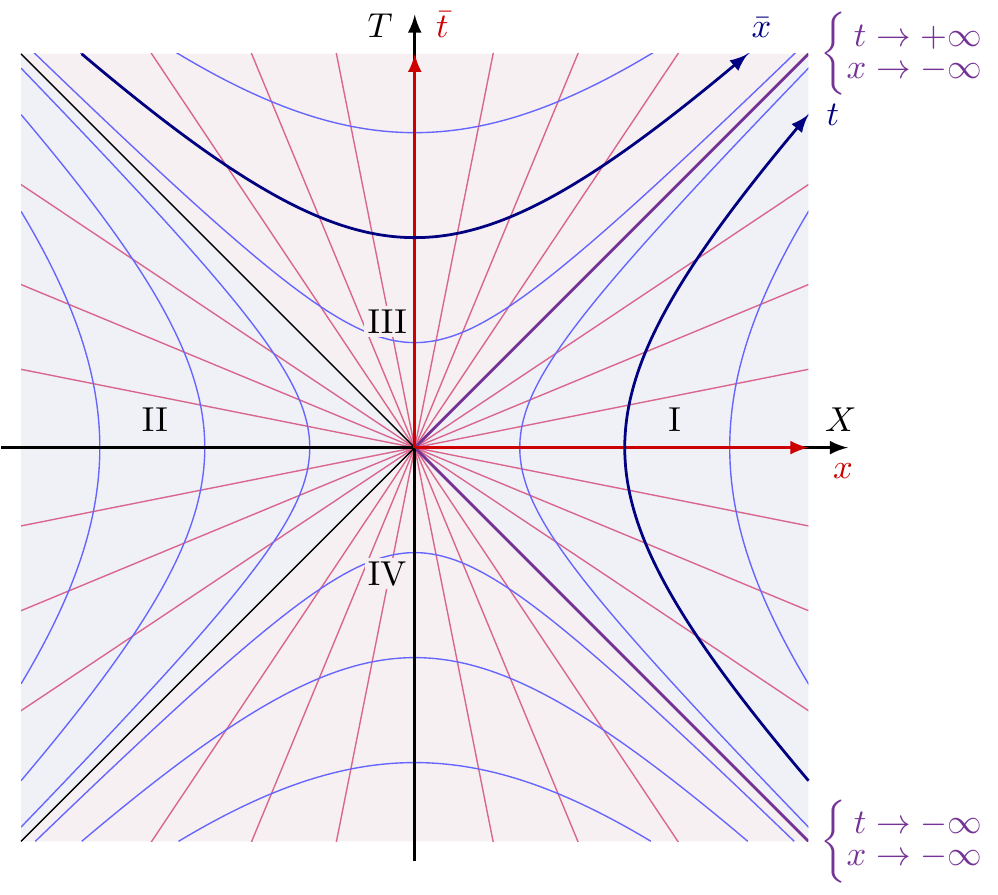}
        \caption{The geometry under consideration via \eqref{Rmetric1} and \eqref{RinMin1} is presented. The Minkowski spacetime \eqref{RinMin1} is covered by $(T,X)$ without any restrictions in the coordinates. On the other hand, the Rindler spacetime \eqref{Rmetric1} covers only the right or left Rindler wedge(region I or II) in $(T,X)$-coordinates, while $-\infty<t,x<\infty$. }
        \label{fig:geo}
        \end{center}
\end{figure}

\begin{figure}[h]
        \begin{center}
        \includegraphics{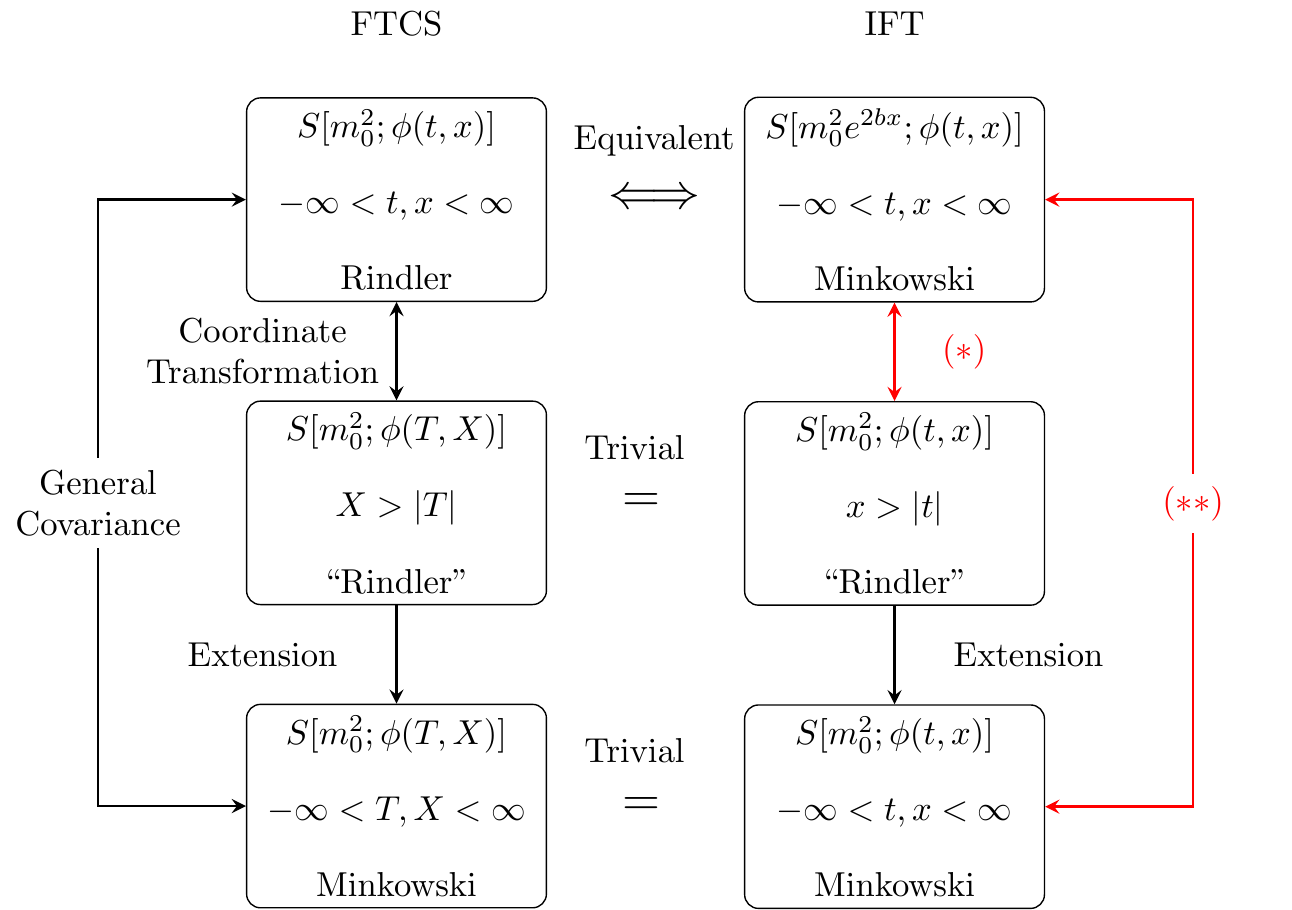}
        \caption{FTCS's and IFT's are presented on left-side and right-side columns, respectively. Our classical equivalence relates FTCS and IFT in the first row. In the left-hand side, the first and the second theories are connected by the coordinate transformation~\eqref{CoordTr}. On the other hand, the theories in the second row are trivially identical to each other as they are nothing more than changing dummy variables on the same region. So the relation $\textcolor{red}{(*)}$ between two different IFT's is deduced.  In the left-hand side, the first and the third theories are connected by general covariance which includes the extension of the range of domains. IFT's with different mass functions in the same range can be related by a hidden relation $\textcolor{red}{(**)}$.
        }
        \label{fig:rel}
        \end{center}
\end{figure}

In order to show this  phenomenon explicitly, we consider the Rindler spacetime and a coordinate transformation in the side of FTCS. First, we take the metric  in the FTCS action as
\begin{align} \label{Rmetric1}
ds^{2} = e^{2bx}(-dt^{2} + dx^{2})\,,
\end{align} 
where $b>0$ corresponds to  a constant proper acceleration, related to a static observer in the Rindler spacetime.
Then the  corresponding IFT action by our equivalence becomes
\begin{align} \label{SIFT1}
S_{{\rm IFT}} = \int d^{2}x \Big[ -\frac{1}{2}\partial _{\mu}\phi \partial^{\mu}\phi -\frac{1}{2}m^{2}(x)\,  \phi^{2}\Big]\,,\qquad m^{2}(x) = m^{2}_{0}\, e^{2bx}\,.
\end{align}
Second, let us carry out  the coordinate transformation   in the FTCS side   as $(t,x)\to (T,X)$, 
\begin{align}\label{CoordTr}
T =  b^{-1} e^{b x} \sinh (b t),\qquad  X =  b^{-1} e^{b x} \cosh (b t)\,.
\end{align}
This brings the Rindler metric \eqref{Rmetric1}  to the Minkowski metric,
\begin{align} \label{RinMin1}
ds^2 = - dT^2 + dX^2
\end{align}
with the range of coordinates 
$X > |T|$. See Fig.~\ref{fig:geo}. This is the so-called right Rindler wedge of the Minkowski spacetime.  This right Rindler wedge spacetime is a globally hyperbolic but geodesically incomplete spacetime. To avoid the encounter with the spacetime boundary within a finite proper time/affine parameter, one usually extend the spacetime to be geodesically complete. In this case, the extended one is the Minkowski spacetime with $-\infty<T,X<\infty$.
This is  natural    in the context of general covariance in FTCS, which is displayed in the left side of Fig.~\ref{fig:rel}. 

%

Now, we convert the transformed FTCS action to the corresponding IFT action by our equivalence, to see the effect of the coordinate transformation in the IFT point of view.  Adopting the equivalence \eqref{IFT=g2}, we obtain the converted action with the full range of coordinates, $T$ and $X$, 
\begin{align} \label{SIFT2}
S_{{\rm IFT}} = \int d^{2}X \Big[ -\frac{1}{2} \bar \partial _{\mu}\phi \bar \partial^{\mu}\phi -\frac{1}{2}\bar{m}^{2} (X)\,  \phi^{2}\Big]\,, \qquad \bar{m}^{2}(X) = m_0^{2}\,,
\end{align}
where $\bar\partial_{\mu} \equiv \frac{\partial}{\partial X^\mu}$.
Superficially, two IFT actions in \eqref{SIFT1} and \eqref{SIFT2} look different.
Nevertheless, any physical quantities derived from $S_{\rm IFT}[m_0^2e^{2bx};\phi(t,x)]$ and $S_{\rm IFT}[m_0^2;\phi(T,X)]$ should be related by a definite connecting rule, according to our equivalence and general covariance of FTCS.
This leads to an unexpected consequence from the perspective in IFT, which is a definite relation\footnote{From the perspective of IFT, two IFTS's of the exponential mass function and  the constant mass can be understood as related by a Weyl rescaling of the flat metric $\eta$ as $\eta \rightarrow e^{2\omega(x)}\eta$, which is not a symmetry of the system.} between the scalar IFT with the exponential mass function and the scalar IFT with constant mass, as displayed in the right side of Fig.~\ref{fig:rel}.

\subsection{Limiting behaviors of mass function}

In this subsection, we investigate limiting behaviors of the mass function $m(x)$ in terms of background metric in FTCS  through our equivalence. In the conventional field theory with a constant mass, $m_0^2$ should be taken as a nonnegative value for a nontachyonic behavior. To avoid unnecessary intricacy we focus on $m^2(x)\geq0$. Under this condition, limiting behaviors of mass function such as $m^{2}(x)\rightarrow0$ and $m^2(x)\rightarrow\infty$ would be interesting. In the case of $m^2(x)=0$, scaling symmetry appears in IFT. This symmetry corresponds to the Weyl symmetry in FTCS. 

Now, we provide a geometrical interpretation  of  massless and infinitely-massive points of the mass function $m(x)$ in the scalar IFT for appropriate $x$-values. 
First of all, the massless  condition  ($m^{2}(x_{*})=0$) can be read from (\ref{Ourmet0}) and (\ref{Matrel})  as 
\begin{equation} \label{massless}
m^{2}(x) \Big|_{x=x_{*}}= m^{2}_{0}\, e^{2\omega(x)} -2\xi \omega''(x)\Big|_{x=x_{*}}=0\,.
\end{equation}
There are two possibilities to achieve this condition, as far as $m_{0}\xi \neq 0$. One possibility is that neither $e^{2\omega(x)}$ nor $\omega''(x)$ vanishes at $x=x_{*}$. 
In this case, $\omega(x)$ should behave as $\omega''(x)  \sim e^{2\omega(x)}$, to satisfy the massless condition. By solving (\ref{massless}), we see that $\omega(x)$ near the massless point $x=x_*$ behaves as 
\begin{equation} \label{localdS}
e^{\omega (x) } \sim \frac{1}{\cosh \gamma (x+c)}\,, \qquad R\sim 2\gamma^{2}\,, \qquad m(x) \sim \frac{\#}{\cosh \gamma (x+c)}\,,  
\end{equation}
or
\begin{equation} \label{localAdS}
e^{\omega (x) } \sim \frac{1}{\sinh \gamma (x+c)}\,, \qquad R\sim -2\gamma^{2}\,, \qquad m(x) \sim \frac{\#}{\sinh \gamma (x+c)}\,,   
\end{equation}
where $\gamma$,  $c$ and $\#$ are some constants. The massless point is given by $|x_*|\rightarrow \infty$. Without loss of generality, we set $c=0$ translating the $x$-variable in our metric. Since the local curvature is constant in these cases, one can identify these special cases as locally dS or flat, and locally AdS spaces. 

The other possibility is that both $e^{2\omega(x)}$ and $\omega''(x)$ independently vanish  at the massless point. 
Note that this point corresponds to (at least) a coordinate singularity since our metric becomes degenerate at this point as can be seen from our metric in (\ref{Ourmet0}). Recalling that the Ricci scalar for our metric is given by ${\cal R} = -2\omega''(x) e^{-2\omega(x)}$,
we see that the Ricci scalar divergence at the massless point is determined by the scaling behaviors of $e^{2\omega(x)}$ and $\omega''(x)$. If the scaling behaviors at the massless point are given by 
\begin{equation} \label{SndPos}
e^{-2\omega(x)}\sim \frac{1}{\epsilon^{2}}\,, \qquad \omega''(x) \sim \epsilon^{2-\alpha}\,, \qquad 0<\alpha<2\,, \quad   \epsilon \rightarrow 0\,,
\end{equation}
the Ricci scalar diverges at the massless point and thus this  corresponds to a curvature singularity\footnote{This second possibility will be studied in the subsequent work~\cite{Ho:2022omx}.}. 

The above observations tell us that the massless point should correspond to either a coordinate singularity or a curvature singularity. Our following examples are not included in the case of (\ref{SndPos})   and we restrict our attention to the first possibility.
In fact, one can show that the coordinate singularity in our cases corresponds to a Killing horizon. 
 This is shown by using a time-like Killing vector, $K=\frac{\partial}{\partial t}$, which always exists for our metric in Eq.~(\ref{Ourmet0}) and whose normalization is taken as $K^{2}\rightarrow -1$ when $e^{2\omega}\rightarrow 1$. In the case of asymptotically AdS or dS spacetime, the normalization of $K$ should be taken appropriately\footnote{The AdS and dS cases are presented in the following.}.
Then, from the definition of a Killing horizon ${\cal H}$, one can see that it is specified by the property $K^{2}|_{\cal H}=0$ which corresponds to $e^{2\omega}=0$. 
Even in the locally dS (\ref{localdS}) and the locally AdS (\ref{localAdS}) spacetimes, which correspond to the static patch of dS spacetime and the Rinder patch of AdS spacetime, respectively, there are Killing horizons specified by $e^{2\omega}=0$ . This tells us that the coordinate singularity given by $e^{2\omega}=0$ always corresponds to a Killing horizon in our case.

On the contrary,  it is rather simple to describe  
the infinitely massive point $m^{2}(x)\rightarrow \infty$
in the corresponding FTCS. The mass formula in~(\ref{Matrel}) tells us that $e^{2\omega}$ or $\omega''$ should diverge at that point. Just like in the massless point case, we focus on  the scaling behaviors of $e^{2\omega}$ and $\omega''$ as 
\begin{equation} \label{}
e^{2\omega} \sim \frac{1}{\varepsilon^{2}}\,,  \qquad  \omega''  \sim \varepsilon^{\beta -2}\,, \qquad   0 <  \beta \,, \quad   \varepsilon \rightarrow 0\,,
\end{equation}
which means that $\omega''$ diverges slowly than $e^{2\omega}$. In this case, the Ricci scalar vanishes at the infinitely massive point.  Then, the infinitely massive point is realized as the flat background metric on the corresponding FTCS.

In the following we present various  examples which provide some concrete realization of scaling behaviors of $e^{2\omega}$ and $\omega''$. As a first example, we consider the Rindler spacetime which is described by $\omega(x) = bx$. See (\ref{SIFT1}). 
In this case, the massless point is given by  $x\rightarrow -\infty$ which  corresponds to the Rindler horizon, while the infinitely massive point  is given by $x\rightarrow \infty$ which corresponds to the future and past null infinities of the Rindler spacetime. 
This example is consistent with our scaling arguments in the above.

In the next we show several non-flat examples.
In the case of dS$_{2}$, the Ricci scalar is given by  $\mathcal{R}= 2/\ell^{2}$ with the de Sitter radius $\ell$. One can see that
\begin{equation} \label{}
m^{2}(x) = \Big( m^{2}_{0} + \frac{2\xi}{\ell^{2}} \Big)\frac{1}{\cosh^{2}(x/\ell)}\,,
\end{equation}
which leads to positive $m(x)$ within our choice $\xi>0$. The massless condition $m(x)\rightarrow0$ is achieved by $x\rightarrow\pm\infty$ which are de Sitter horizons in our metric.
In the case of AdS$_{2}$, the Ricci scalar is given by $\mathcal{R}= -2/\ell^{2}$ with the anti-de Sitter radius $\ell$. Here we take the range of $x$ as $(0,+\infty)$. In this case,  $x=0$ corresponds to the boundary of AdS$_{2}$ and 
\begin{equation} \label{}
m^{2}(x) = \Big( m^{2}_{0} -  \frac{2\xi}{\ell^{2}} \Big)\frac{1}{\sinh^{2}(x/\ell)}\,,
\end{equation}
where $m(x)$ always becomes positive under our choice $\xi < 0$ and 
\begin{equation} \label{}
m(x) \rightarrow \infty  \quad \text{as} \quad x\rightarrow 0\,, \qquad m(x) \rightarrow  0  \quad \text{as} \quad x\rightarrow \infty\,.
\end{equation}
Indeed the massless point and the infinitely massive point correspond to the Killing horizon and the flat region, respectively.

Let us consider $(1+1)$-dimensional black hole backgrounds, whose metric is taken as~\cite{Mandal:1991tz,Giddings:1992ff}
\begin{equation} \label{}
ds^{2} = -\frac{dudv}{ 1 - \lambda^{2}uv}\,.
\end{equation}
Under the coordinate transformation
\begin{align}
u = -\frac{1}{b}e^{-b(t-x)}\,, \qquad v = \frac{1}{b}e^{b(t+x)} \,,
\end{align}
we can set it as
\begin{equation} \label{MSWmetric}
ds^{2} = \frac{-dt^2+dx^2}{\frac{\lambda^{2}}{b^2} + e^{-2bx} }\,.
\end{equation}
Then the mass function in IFT is given by
\begin{equation} \label{}
m^{2}(x) = \Big( m^{2}_{0} + \frac{4\xi\lambda^2e^{-2bx}}{\frac{\lambda^2}{b^2}+e^{-2bx}} \Big)\frac{1}{\frac{\lambda^2}{b^2}+e^{-2bx}}\,,
\end{equation}
which is positive for $\xi >0$ and 
\begin{equation}  
m(x) \longrightarrow %
\left\{ \begin{array}{lll}    
m_{0}/\lambda&   \text{as} & x\rightarrow +\infty  \\
~ 0&     \text{as} & x\rightarrow -\infty 
\end{array}  \right.\,.   \label{}
\end{equation}
So the massless point corresponds to the black hole horizon and the infinitely massive point does not exist. The limit of $\lambda\rightarrow1$ is nothing but the Witten's $(1+1)$-dimensional black hole~\cite{Witten:1991yr}. Note also that in the limit of $\lambda\rightarrow0$, the metric becomes the Rindler spacetime. 

In all the above examples, massless points and infinitely massive points in IFT correspond to Killing horizons and flat regions in FTCS, respectively. Note that the scaling argument is valid for all the above examples.

\section{Quantization}\label{QIFT}

In the previous section, we have discussed the classical equivalence between the scalar IFT and the scalar FTCS in $(1+1)$ dimensions. Based on this equivalence, we would like to give a proposal for the quantization of IFT. 
Our proposal includes the canonical quantization for the specific IFT, which is read from the canonical quantization procedure in the FTCS (See  Appendix~\ref{AppB}). This proposal tells us how to quantize the specific IFT. As an example, we consider the quantization of the specific IFT in which the mass function is given in an exponential form.
Yet, it has been known that the canonical quantization of FTCS is insufficient to provide a comprehensive general framework, at least conceptually. A more adequate and conceptually favorable framework is based on an algebraic construction, which is called an algebraic formulation of {\it quantum field theory on curved spacetime} (QFTCS).  Our proposal actually is made on this framework. Some aspects in this framework are given in the next section.

\subsection{Proposal}
The quantization of IFT is not well established in a comprehensive way contrary to the conventional homogeneous field theory which possesses the Poincar\'{e} symmetry. Concretely speaking, the Poincar\'{e} invariant vacuum does not exist because the position-dependent mass and couplings break the Poincar\'{e} symmetry explicitly. This situation is somewhat similar to the case of FTCS in which the preferred vacuum state cannot be selected because of the absence of the Poincar\'{e} symmetry in a generic curved spacetime background. This naturally motivates us to adopt similar quantization method in the FTCS to the IFT. Triggered by this motivation and elevating the classical equivalence to the quantum level, we arrive at the following proposal to the quantization of IFT in $(1+1)$ dimensions:
\begin{equation}
        \boxed{\text{IQFT $\iff$ QFTCS}}\,.
\end{equation}
One implication of this proposal is that the canonical quantization in FTCS is transcribed to that of IFT. As a concrete example, we will consider the field theory on the Rindler spacetime \eqref{Rmetric1} and the corresponding IFT with exponential mass function in the next subsection. Our proposal on quantum equivalence is made on the class of IFT within the validity of the `classical equivalence'.

This proposal is not just about canonical quantization, but has broader implications.
As is well-known, the difficulty in introducing the preferred vacuum state on  curved spacetime leads to an algebraic formulation of QFTCS. 
 Briefly speaking, the algebraic formulation starts from appropriate algebraic relations among quantum fields (local algebra)  with some appropriate properties or axioms: isotony, covariance, locality/causality, and the existence of dynamics (also known as the time slice axiom).  And then, algebraic states are introduced as normalized positive linear functionals on the field algebra. Through the so-called Gelfand-Naimark-Segal (GNS) construction, we can construct a relevant Hilbert space from the algebraic states. Especially, the free field case can be formulated in a rigorous way.  See~\cite{Haag:1992hx,Wald:1995yp,Yngvason:2004uh,Halvorson:2006wj,Hollands:2009bke,Benini:2013fia,Hollands:2014eia,Khavkine:2014mta,Fredenhagen:2014lda} for reviews on algebraic formulation of QFTCS.
Our proposal incorporates these aspects of QFTCS into the quantum equivalence between the scalar IQFT and the scalar QFTCS.

We would like to emphasize that the classical equivalence in the previous section does not automatically warrant its quantum version. As is well-known, the ordering ambiguity in the operator elevation of classical variables results in the trivial example of inequivalent quantum theories with the classical equivalence. Therefore, our proposal should be taken as one possible way to quantize IFT and may be tested only by experiments.

Of course, this quantum equivalence should be taken with some caution, since a generic IFT cannot be realized even classically by a FTCS in $(1+1)$ dimensions. Therefore, the above quantum equivalence should be taken only when the classical equivalence holds. Though our proposal, which is applicable only in the form of metric \eqref{Ourmet0}, is not completely generic, it leads to a concrete way to compute the quantum effects in the IFT. In the following, we present a canonical quantization method in a specific IFT model before giving an algebraic Hadamard approach in the next section.

\subsection{IQFT with an exponential mass function}
According to our proposal, the quantization of IFT with an exponential mass function is achieved by the canonical quantization on the Rindler spacetime. First we present a brief summary of this procedure. The Klein-Gordon equation of the IFT with the mass function $m(x)=m_0e^{bx}$ is given by
\begin{equation}
\left(\frac{\partial^2}{\partial t^2}-\frac{\partial^2}{\partial x^2} +m_0^2e^{2bx} \right)\phi(t,x)=0\,,
\end{equation}
which is the same form for the scalar field in the Rindler spacetime. This result is a trivial consequence from our equivalence. Thus we can apply all the results in the Rindler spacetime to the IFT with the exponential mass function. In particular, the mode solution of the equation of motion on the Rindler spacetime is given by~\cite{Fulling:1972md,Takagi:1986kn,Fulling:1989nb}
\begin{align}
        &u_{\Omega}(t,x)=\frac{1}{\sqrt{2\Omega}}\theta(\rho)h_{\Omega}(\rho)e^{-i\Omega \eta}\,,
        \\
        &\eta\equiv bt\,,\quad \rho\equiv b^{-1}e^{bx}\,,\label{rhoeta}
\end{align}
where $\Omega$ is a positive energy eigenvalue of the mode, $\theta$ is the step function, and $h$ is given by the modified Bessel function of second kind as
\begin{equation}
        h_{\Omega}(\rho)=R^*\sqrt{\frac{2}{\pi}}\frac{K_{i\Omega}(m_0\rho)}{|\Gamma(i\Omega)|}\,,\qquad R^*\equiv \left(\frac{m_0}{2b}\right)^{2i\Omega} \left(\frac{\Gamma(-i\Omega)}{|\Gamma(i\Omega)|}\right)^2\,.
\end{equation}
These mode solutions can be used for the IFT. 

As is well-known, it is straightforward to perform the canonical quantization in the Rindler spacetime. A brief explanation of the canonical quantization on  curved spacetime is given in Appendix \ref{AppB}. The equal time commutator
\begin{align}
&[ \phi(t,x), \pi (t, y) ] = i\delta (x-y)\,
\end{align}
gives the commutation relation $[b_{\Omega},b^{\dagger}_{\Omega'}]=\delta(\Omega-\Omega')$, where $b_{\Omega}$ and $b^{\dagger}_{\Omega}$ are the annihilation and the creation operators, respectively. And then the scalar field is expanded as:
\begin{equation}
        \phi(t,x)=\int_{0}^{\infty}d\Omega\,\Big(b_{\Omega}\,u_{\Omega}(t,x)+b_{\Omega}^{\dagger}\,u_{\Omega}^*(t,x)\Big)\,.
\end{equation}
The Rindler vacuum, $|0\rangle_{\rm R}$, is defined by
\begin{equation}
        b_{\Omega}|0\rangle_{\rm R}=0\,,
\end{equation}
from which we can construct the Fock space with creation operators, $b^{\dagger}_{\Omega}$.

All these results in  QFTCS are transcribed to  IQFT setup. According to our quantum equivalence, the Rindler vacuum, $|0\rangle_{\rm R}$, and the annihilation/creation operators, $b_{\Omega}/b_{\Omega}^{\dagger}$, are identified with those of  IQFT.
So the Fock spaces of  QFTCS and  IQFT are identical and so does the two-point functions. Since we are considering free field theories, the two-point functions of both sides determine any $n$-point functions. Therefore all $n$-point functions of two theories are the same, which is the meaning of our quantum equivalence.  To emphasize that the quantization is done in the context of the scalar IFT, we denote the vacuum of the scalar IQFT with an exponential mass function as $|\underline{0}\rangle_{\rm IFT}$.

In the next section, we explore further aspects of the quantum equivalence focusing on the Unruh effect.

\section{Quantum Aspects of IQFT}
In this section, we present a quantization of  IFT based on the algebraic method which supersedes the canonical quantization considered in the previous section. One feature of the algebraic approach is that it democratically treats the pure and the mixed states in the canonical quantization.
From our proposal that the $(1+1)$-dimensional IQFT is equivalent to the $(1+1)$ dimensional QFTCS, it is natural to anticipate that the quantization and renormalization methods in  QFTCS should be carried over to IQFT in a rather straightforward manner. After a brief review on the Hadamard method in  QFTCS, we apply this to  IQFT to see some quantum effects including the Unruh effect.

\subsection{Hadamard renormalization}
  In the algebraic formulation of QFTCS, a physically important class of quantum states are given by Gaussian Hadamard states which may serve as substitute of the preferred vacuum state.  Hadamard state is defined as  an algebraic state satisfying the Hadamard condition which is motivated by some reasonable physical considerations. The condition includes that the short distance singularity structure of the $n$-point functions of the Hadamard state on curved spacetime should be given by that of the $n$-point functions of the vacuum state in the Minkowski spacetime, the ultra-high energy mode of quantum fields resides essentially in the ground state, and the singular structure of the $n$-point functions should be of positive frequency type~\cite{Fulling:1978ht,Fulling:1981cf,Kay:1988mu,Radzikowski:1996pa}. 
  
  Contrary to the ordinary vacuum state, the Hadamard state is not unique for a given background spacetime but forms a class in general. In the case of a Gaussian Hadamard state\footnote{A Gaussian state, which is also called a quasi-free state, is defined by the condition that the connected $n$-point functions of the state vanish, or any  $n$-point functions can be obtained from $1$- and $2$-point functions.}, one can obtain a Fock space representation of the algebra of  quantum fields and can identify the Gaussian Hadamard state with the Fock space vacuum. In this way, one can see that some well-known vacua belong to the Hadamard class. The Hadamard method encompasses the usual Fock space canonical quantization and implements appropriately relevant requirements such as general covariance of stress tensor, while it connects unitarily inequivalent representations of the algebras of observables.~\cite{Kay:1988mu,Wald:1995yp,Hollands:2014eia}.

Under this scheme, the Gaussian Hadamard state, $\omega_{\rm H}$ is defined by the renormalized two point function of scalar field $\phi$ as\footnote{The normalization condition of an algebraic state $\omega$, is $\omega(\mathbf{1})=1$, where $\mathbf{1}$ is an identity element in the field algebra. }
\begin{equation} \label{HadamardFTCS}
\omega_{\rm H}\big(\phi (\mathbf{x}) \phi (\mathbf{x}')\big) =  F(\mathbf{x},\mathbf{x}') - H(\mathbf{x},\mathbf{x}') \,,  
\end{equation}
where $F(\mathbf{x},\mathbf{x}')$ is an unrenormalized two point function known as the Hadamard function, and    the function $H(\mathbf{x},\mathbf{x}')$ is so called as the Hadamard parametrix~\cite{Hadamard:1952}. In the following, $\omega_{\rm H}(\phi (\mathbf{x}) \phi (\mathbf{x}'))$ is also denoted by $\langle \phi (\mathbf{x}) \phi (\mathbf{x}')\rangle_{\rm H} $.  Here, $H(\mathbf{x},\mathbf{x}')$  is a local  covariant function of the half of squared geodesic length, $\sigma(\mathbf{x},\mathbf{x}')$ between two points $\mathbf{x}$ and $\mathbf{x}'$, written in terms of the metric and the curvature. The Hadamard function $F(\mathbf{x},\mathbf{x}')$ is symmetric and satisfies
\begin{equation} \label{kgeq}
(-\Box +m^{2} + \xi {\cal R}) F(\mathbf{x},\mathbf{x}') = \delta (\mathbf{x}-\mathbf{x}')\,,
\end{equation}
which can also be represented by a real part of the `positive frequency' $2$-point Wightman function, $\textrm{Re}~ G^{+}(\mathbf{x},\mathbf{x}')$. It is known that $H(\mathbf{x},\mathbf{x}')$ takes the same form for any Hadamard states. Then, the Hadamard renormalization is achieved by  subtracting the singular part $H(\mathbf{x},\mathbf{x}')$ from the function $F(\mathbf{x},\mathbf{x}')$.

Explicitly, the Hadamard parametrix $H$ is given by  
\begin{align} \label{}
H(\mathbf{x},\mathbf{x}') &= \alpha_{D} \frac{U(\mathbf{x},\mathbf{x}')}{\sigma^{\frac{D}{2}-1}(\mathbf{x},\mathbf{x}')} + \beta_{D} V(\mathbf{x},\mathbf{x}') \ln \left(\mu^2\, \sigma(\mathbf{x},\mathbf{x}')\right) \quad\text{for even $D$}\,,\nonumber \\
H(\mathbf{x},\mathbf{x}') &= \alpha_{D} \frac{U(\mathbf{x},\mathbf{x}')}{\sigma^{\frac{D}{2}-1}(\mathbf{x},\mathbf{x}')} \quad\hskip4.8cm \text{for odd $D$}\,,
\end{align}
where $\alpha_{D}, \beta_{D}$ are numerical constants depending on the dimension $D$ and  $\mu$ is a certain mass scale introduced from the dimensional reason. Symmetric bi-scalars $U(\mathbf{x},\mathbf{x}')$ and $V(\mathbf{x},\mathbf{x}')$, which are regular for $\mathbf{x}' \rightarrow \mathbf{x}$, are universal geometrical objects independent of any Hadamard states. They can be expanded in terms of $\sigma (\mathbf{x}, \mathbf{x}')$ as
\begin{equation} \label{}
U(\mathbf{x},\mathbf{x}')  =  \sum_{n=0}^{D/2 - 2} U_n (\mathbf{x},\mathbf{x}') \sigma^n (\mathbf{x}, \mathbf{x}') , \qquad  V(\mathbf{x},\mathbf{x}')  =  \sum_{n=0}^{+ \infty} V_n (\mathbf{x},\mathbf{x}') \sigma^n (\mathbf{x}, \mathbf{x}') .
\end{equation}
Here, $U_n (\mathbf{x},\mathbf{x}')$ and $V_n (\mathbf{x},\mathbf{x}')$ can be completely determined by recursion relations and boundary conditions, which are obtained by comparing the power of $\sigma$ on the both sides of the equation, $(-\Box +m^{2} + \xi {\cal R}) H(\mathbf{x},\mathbf{x}') = \delta (\mathbf{x}-\mathbf{x}') $ ~\cite{Wald:1977up}.
Concretely,  in two dimensions $\alpha_2=0$ and the Hadamard parametrix becomes
\begin{equation}
H(\mathbf{x},\mathbf{x}') = \frac{V(\mathbf{x},\mathbf{x}')}{2\pi} \ln \left(\mu^2\, \sigma(\mathbf{x},\mathbf{x}') \right)\,,
\end{equation}
where the bi-scalar $V(\mathbf{x},\mathbf{x}') $ is given by~\cite{Decanini:2005eg}
\begin{equation}
V(\mathbf{x},\mathbf{x}') = -1 - \frac{1}{24} \mathcal{R} g_{\mu \nu} \nabla^{\mu} \sigma \nabla^{\nu} \sigma - \frac{1}{2} \left( m_0^2 + \xi - \frac{1}{6}  \right) \mathcal{R} \sigma + {\cal O}(\sigma^{3/2}).
\end{equation}

Since the renormalized stress tensor enters in various kinds of semi-classically improved energy conditions, it has been one of  important topics in QFTCS.
Based on the above procedure, one can obtain the renormalized stress tensor by acting an appropriate differential bi-vector operator, ${\cal T}_{\mu\nu'}$ on the renormalized $2$-point function as
\begin{equation} \label{StressTensor}
\langle T_{\mu\nu} (\mathbf{x}) \rangle_{\rm H} = \lim_{\mathbf{x}'\rightarrow \mathbf{x}}  {\cal T}_{\mu\nu'} \langle \phi (\mathbf{x}) \phi (\mathbf{x}')\rangle_{\rm H}\,.
\end{equation}
For instance, in our two dimensional case~\eqref{GraLag2}, the differential bi-vector is given by
\begin{align}    \label{DiffBiV}
{\cal T}_{\mu\nu'}   = & (1-2\xi) \partial_{\mu}\partial_{\nu'}  +\Big(2\xi- \frac{1}{2}\Big)g_{\mu\nu'} g^{\alpha\beta'}\partial_{\alpha}\partial_{\beta'}    -\frac{1}{2}g_{\mu\nu'}m_0^{2}   \nonumber \\
  &  -2\xi\delta^{\mu'}_{\mu}\partial_{\mu'}\partial_{\nu'} + 2\xi g_{\mu\nu'}g^{\alpha\beta}\nabla_{\alpha}\nabla_{\beta}   +\xi \Big(\mathcal{R}_{\mu\nu'} - \frac{1}{2} g_{\mu\nu'}\mathcal{R} \Big)\,.
\end{align}
 Though there still remain some ambiguities in this subtraction method, one can construct an essentially unique stress tensor under Wald's axiom. The ambiguous terms in stress tensor are written in geometrical quantities~\cite{Wald:1995yp,Moretti:2001qh}. 
Sometimes, the above procedure has been known as a `point-splitting method'~\cite{Birrell:1982ix}, while it is now regarded as more reliable results on a firm mathematical ground. We will apply this well-established prescription in  QFTCS to  IQFT in the following subsection.

Some comments are in order~\cite{wtkim}.
The stress tensor in QFTCS is expected to satisfy some natural axioms~\cite{Wald:1978pj}. 
For instance, the quantum expectation value of the stress tensor should be local, covariant, covariantly-conserved, and etc.
Now, it is widely believed that the so-called Hadamard renormalization is well suited to this purpose. For instance, a global Hadamard state consistent with the Hadamard renormalization would lead to 
\begin{equation} \label{Hadamard}
\langle T_{\mu\nu} (\mathbf{x}) \rangle_{\rm H} = \frac{\partial y^{\alpha}}{\partial x^{\mu}}  \frac{\partial y^{\beta}}{\partial x^{\nu}} ~ \langle T_{\alpha\beta}(\mathbf{y}) \rangle_{\rm H} \,,
\end{equation}
where it may be noted that the vacuum states in each coordinate $\mathbf{x}$ and $\mathbf{y}$ do not need to be realized on the same Fock space in general. 
Note that the Minkowski vacuum is a global Hadamard state on the whole Minkowski space, while the Rindler vacuum is not a global Hadamard one on the whole Minkowski space since it diverges on the Rindler horizon. However, the Rindler vacuum is a Hadamard state on the Rindler wedge. To proceed, let us denote the Poincar\'{e} invariant Minkowski vacuum $|0\rangle_{\rm M}$ and the Rindler vacuum $|0\rangle_{\rm R}$ on each Fock spaces.  
As is well-known, the Minkowski vacuum is realized as a Kubo-Martin-Schwinger(KMS) state on the Rindler wedge~\cite{Bisognano:1976za,Kay:1985zs}.

To avoid cluttering the discussion, we repeat the expression in \eqref{Hadamard} in terms of the algebraic state $\omega_{\rm M}$ which is a Gaussian Hadamard state in the Minkowski spacetime as
\begin{equation} \label{omegaRel}
\omega_{\rm M}(T^{\rm M}_{\mu\nu}(\mathbf{x})) = \frac{\partial y^{\alpha}}{\partial x^{\mu}}  \frac{\partial y^{\beta}}{\partial x^{\nu}} ~ \omega_{\rm M}( T^{\rm R}_{\alpha\beta}(\mathbf{y}) )\,,
\end{equation}
where $T^{\rm M}_{\mu\nu}(\mathbf{x})$ denotes the stress tensor on the Minkowski spacetime and $T^{\rm R}_{\alpha\beta}(\mathbf{y})$  does from the Rindler one. 
In the Fock space representation of this state on the Minkowski spacetime is give by
\begin{equation} \label{}
\omega_{\rm M}(T^{\rm M}_{\mu\nu}(\mathbf{x}))={}_{\rm M}\langle  0| T^{\rm M}_{\mu\nu} (\mathbf{x})|0  \rangle_{\rm M}\,,
\end{equation}
which is represented by a KMS state on the Rindler patch. In the standard convention, we take the Minkowski vacuum energy to be zero which means that $\omega_{\rm M}(T^{\rm M}_{\mu\nu})=0$.

Now we can consider the Gaussian Hadamard state on the right Rindler wedge, $\underline{\omega}_{\rm R}$ which is not a global Hadamard state on the Minkowski spacetime, since it diverges on the Rindler horizon. Just like the algebraic state $\omega_{\rm M}$, in the Fock space representation of $\underline{\omega}_{\rm R}$ on the Rindler spacetime, $\underline{\omega}_{\rm R}$ is given by the Rindler vacuum, as
\begin{equation} \label{}
\underline{\omega}_{\rm R}(T^{\rm R}_{\mu\nu}(\mathbf{y}))=\langle T_{\mu\nu}^{\rm R}(\mathbf{y}) \rangle_{\rm H}^{\rm R} ={}_{\rm R}\langle  0| T^{\rm R}_{\mu\nu} (\mathbf{y})|0  \rangle_{\rm R}<0\,,
\end{equation}
which  is   the minimum energy ground state in the Rindler wedge by the definition, while the Minkowski vacuum is excited one as $\omega_{\rm M}(T^{\rm R}_{\mu\nu})=0$. Note  that this is consistent   with the Unruh effect. See~\eqref{NegE2}.

Another way to understand the Unruh effect in terms of stress tensor is to consider a normal ordering prescription. Let us define a normal ordering of stress tensor  operator in the Minkowski spacetime by the subtraction of its vacuum expectation value as 
\begin{equation} \label{}
:T^{\rm M}_{\mu\nu}(\mathbf{x}) :  \;\equiv T^{\rm M}_{\mu\nu}(\mathbf{x})  - {}_{\rm M} \langle 0| T^{\rm M}_{\mu\nu}(\mathbf{x}) | 0\rangle_{\rm M}~ {\bf 1}\,, 
\end{equation}
where ${\bf 1}$ denotes the identity operator. Our choice of $\omega_{\rm M}(T^{\rm M}_{\mu\nu})=0$ means $:T^{\rm M}_{\mu\nu}(\mathbf{x}) :=\;T^{\rm M}_{\mu\nu}(\mathbf{x})$.
By taking the same definition of a normal ordering in the Rindler spacetime, 
\begin{equation} \label{}
:T^{\rm R}_{\mu\nu}(\mathbf{y}) : \; = T^{\rm R}_{\mu\nu}(\mathbf{y})  - {}_{\rm R} \langle 0| T^{\rm R}_{\mu\nu}(\mathbf{y}) | 0\rangle_{\rm R}~ {\bf 1}\,, 
\end{equation}
one can see that
\begin{equation} \label{}
\omega_{\rm M}( :T^{\rm R}_{\mu\nu}(\mathbf{y}) : )  = \omega_{\rm M}( T^{\rm R}_{\mu\nu}(\mathbf{y})) - {}_{\rm R} \langle 0| T^{\rm R}_{\mu\nu}(\mathbf{y}) | 0\rangle_{\rm R}= - {}_{\rm R} \langle 0| T^{\rm R}_{\mu\nu}(\mathbf{y}) | 0\rangle_{\rm R} ~ > ~ 0\,, 
\end{equation}
where we used $\omega_{\rm M}(T^{\rm R}_{\mu\nu})=0$. This is another way of explanations for the Unruh effect.

\subsection{Interpretation in  IQFT}
In this subsection, we focus on the IQFT with the exponential mass function, which corresponds to the quantum field theory on the Rindler spacetime background~\eqref{Rmetric1}. According to our quantum equivalence, all the construction of previous subsection can be transcribed to the scalar IQFT. Especially, we consider the stress tensor construction in the scalar IQFT.
The procedure of the construction goes as follows.  The vacuum expectation value of stress tensor in QFTCS is given by (recall that $V(\mathbf{x},\mathbf{x}')=-1$ in the flat case)
\begin{equation} \label{}
\langle T_{\mu\nu}\rangle_{\rm H} = \lim_{\mathbf{x}'\rightarrow \mathbf{x}} {\cal T}_{\mu\nu'} \bigg[ F(\mathbf{x},\mathbf{x}') + \frac{1}{4\pi}\ln \left( \mu^{2}\sigma(\mathbf{x},\mathbf{x}') \right) \bigg]\,.
\end{equation}

We are interested in the counterpart in the scalar IQFT of the Unruh effect in the scalar QFTCS by setting $\xi=0$ in \eqref{GraLag2}.
To simplify the description, we take the limit of $m_0\rightarrow0$. Because of the infra-red divergence, the $(1+1)$-dimensional scalar theory with $m_0=0$ is not well-defined in the strict sense~\cite{Coleman:1973ci}. However, in ~\cite{Takagi:1986kn}, the limit of $m_0\rightarrow0$ is carefully taken into account to see the Unruh effect in the massless case. In the context of our quantum equivalence, all the consequences of  QFTCS including the results and the process of the limit $m_0\rightarrow0$ are carried over to IQFT.

The Hadamard function in the limit of $m_0\rightarrow0$ is given by~\cite{Dowker:1978aza,Moretti:1995fa}
\begin{align} \label{}
F(\mathbf{x},\mathbf{x}') &= \frac{1}{2}\,{}_{\rm R}\langle0|\{\phi(\mathbf{x}),\phi(\mathbf{x}')\}|0\rangle_{\rm R} \nonumber \\
&=  \frac{1}{4\pi}\ln \frac{1}{\rho\rho' |\alpha^{2} -(\eta-\eta')^{2}|}\,, \qquad \cosh \alpha \equiv 1+ \frac{(\rho-\rho')^{2}}{2\rho\rho'}\,,
\end{align}
where $\rho$ and $\eta$ have been introduced in~\eqref{rhoeta}. In the IFT coordinates the Hadamard function becomes
\begin{align}
        F_{\rm IFT}(\mathbf{x},\mathbf{x}')&= \frac{1}{2}\,{}_{\rm IFT}\langle\underline{0}|\{\phi(\mathbf{x}),\phi(\mathbf{x}')\}|\underline{0}\rangle_{\rm IFT} \nonumber \\
        &=-\frac{b}{4\pi}(x+x')-\frac{1}{4\pi}\ln \Big|(x-x')^2-(t-t')^2 \Big|\,,
\end{align}
which can be interpreted as the Hadamard function in the IFT with the exponential mass function.
By using the explicit form of the squared geodesic distance in this case,
\begin{equation}
        2\sigma = \Big| (\rho-\rho')^{2}-\rho\rho'\Big(2\sinh\frac{\eta-\eta'}{2}\Big)^{2} \Big|=\frac{4}{b^2}e^{b(x+x')}\Big|\sinh^2\frac{b}{2}(x-x')-\sinh^2\frac{b}{2}(t-t') \Big|\,,
\end{equation}
one can obtain the following expanded expression in the limit of $\mathbf{x}\rightarrow\mathbf{x}'$,
\begin{equation} \label{}
F_{\rm IFT}(\mathbf{x},\mathbf{x}') +  \frac{1}{4\pi}\ln 2\sigma
=\frac{b^2}{48\pi}\big[(x-x')^2+(t-t')^2\big]+\cdots \,,
\end{equation}
where the renormalization scale $\mu$ is removed because it does not affect our result.

From our quantum equivalence, we anticipate that the Unruh-like effect in the scalar IQFT with the exponential mass function. To see this effect, we apply the Hadamard method borrowed from \eqref{StressTensor} to our case, resulting in
\begin{equation} \label{}
\langle T_{\mu\nu}^{\rm IFT} (\mathbf{x}) \rangle_{\rm H}^{\rm IFT} \equiv \lim_{\mathbf{x}'\rightarrow \mathbf{x}}  {\cal T}^{\rm IFT}_{\mu\nu'} \langle \phi (\mathbf{x}) \phi (\mathbf{x}')\rangle_{\rm H}^{\rm IFT}\,,
\end{equation}
where the Hadamard state in IFT is defined in the same way with FTCS in~\eqref{HadamardFTCS}.
And the differential bi-vector,
\begin{align}    \label{}
{\cal T}_{\mu\nu'}^{\rm IFT}   = &  \partial_{\mu}\partial_{\nu'} -\frac{1}{2} \eta_{\mu\nu'} \eta^{\alpha\beta'}\partial_{\alpha}\partial_{\beta'}    -\frac{1}{2}\eta_{\mu\nu'}m^{2}(x) \,,
\end{align}
comes from the classical expression of the ``stress tensor'' in \eqref{Tclassical} with $\xi=0$. The straightforward computation in  IQFT leads to the vacuum expectation value of ``stress tensor'' in the form of
\begin{equation} \label{NegE1}
        \langle T_{tt}^{\rm IFT}\rangle_{\rm H}^{\rm IFT}=\langle T_{xx}^{\rm IFT}\rangle_{\rm H}^{\rm IFT}=-\frac{b^2}{24\pi}\,,\quad \langle T_{\rm IFT}{}^{\mu}_{~\mu} \rangle_{\rm H}^{\rm IFT} =0\,.
\end{equation}
This result is the counterpart of the well-known result~\cite{Unruh:1976db,Takagi:1986kn,Crispino:2007eb,Fulling:2018lez} for the Unruh effects for minimally coupled massless scalar field in $(1+1)$-dimensional Rindler spacetime background,
\begin{equation} \label{NegE2}
\langle T_{\eta\eta}^{\rm R}\rangle_{\rm H}^{\rm R} = -\frac{1}{24\pi}\,, \qquad \langle T_{\rho\rho}^{\rm R}\rangle_{\rm H}^{\rm R} = -\frac{1}{24\pi}\frac{1}{\rho^{2}}\,, \qquad \langle T_{\rm R}{}^{\mu}_{~\mu} \rangle_{\rm H}^{\rm R} =0\,. 
\end{equation}
Indeed by the coordinate transformation in~\eqref{rhoeta}, \eqref{NegE2} is covariantly transformed to \eqref{NegE1}.
Note that the Rindler vacuum energy is negative relative to the Minkowski vacuum energy. So it is not strange that the energy of the vacuum in the IQFT with the exponential mass function is negative in \eqref{NegE1}.

All the above results can be translated to the Unruh-like effect in  IQFT. Therefore we anticipate an experimental verification of the Unruh effect in the setup of  IQFT. There are many attempts to capture the Unruh effect by using hydrodynamical analog of the Schwarzschild metric~\cite{Unruh:1981,Chen:1998kp} or in high energy experiments~\cite{Lynch:2019hmk} which have many technical hurdles to overcome. If one can engineer a $(1+1)$-dimensional condensed matter system realizing the scalar IFT with an exponential mass function, it would be easier to verify the Unruh effect experimentally.

\section{Conclusion}

IFT does not have the Poincar\'{e} symmetry and so it disallows the conventional quantization method. In order to overcome such difficulties, we suggested a kind of dual description, which allows a quantization of IFT. As a first step toward quantization of IFT, we concentrated on the classical equivalence between a scalar FTCS and its corresponding scalar IFT through the explicit expressions of actions and the equations of motion in $(1+1)$-dimensions. 
Along this line, we proposed a generalized ``stress tensor'' in IFT motivated from the counterpart in FTCS. This ``stress tensor'' is conserved by using a covariant derivative newly introduced in IFT. 
Within a free scalar IFT, only the meaningful parameter function in the action is the mass function. Thus some details about limiting behaviors of mass functions in IFT are explored.
We have shown that the massless point of the mass function in IFT corresponds to the horizon of the background spacetime in FTCS.
 In this regard, we expect that physical properties on the horizon would be related to those of massless point of $(1+1)$-dimensional scalar IFT.
As is well-known, FTCS enjoys general covariance. When the classical equivalence is combined with this general covariance, an interesting connection among IFT's is obtained. This connection is given by the procedure of ``equivalence --- general covariance --- equivalence''. See Fig.~\ref{fig:rel}.

Based on the classical equivalence, we proposed a quantum equivalence of QFTCS and IQFT. As an example, we have studied IQFT with an exponential mass function which is shown to be equivalent with quantum field theory on  Rindler spacetime.
Especially, we have identified the Unruh-like effect in IQFT.
Along this line, it is natural to consider other patch of  Minkowski spacetime shown in region III in Fig.~\ref{fig:geo}. We can take a coordinate transformation in this patch as
\begin{align}
        T-X=\frac{1}{b}e^{b(\bar{t}-\bar{x})}\,,\qquad T+X=\frac{1}{b}e^{b(\bar{t}+\bar{x})}\,.
\end{align}
In this case the mass function depends on the time-like coordinate $\bar{t}$: $m^2(\bar{t})=m_0^2e^{2b\bar{t}}$. See Fig.~\ref{fig:geo}. Therefore two differently looking IFTs with mass functions, $m(x)$ and $m(\bar{t})$, are related by general covariance. It would be interesting to study the relation between two IFTs at quantum level. See also~\cite{Akhmedov:2021agm}.

There are many open issues related to our work to be pursued. We expect that our equivalence for scalar field theories can be extended to other field theories: fermion, gauge, tensor, and higher spin ones. Thus the extension to supersymmetric field theory would be possible. The extension to a higher dimensional case is another important future direction. It would also be interesting to explore finite temperature effects in IQFT. Another interesting direction is to include interactions in IFT. In conjunction with condensed matter physics, it is desirable to study non-relativistic limit of IQFT. One interesting subject is to implement IQFT in the context of AdS/CFT correspondence. Since the ``connection path'' of IFT's given by ``equivalence --- general covariance --- equivalence'' may not be unique, the (in)dependence of the path from the view point of IFT needs to be studied further. 

From a more provisional point of view, it would be interesting to identify the type of von Neumann algebra factors  for a local algebra of IFT, which is related to a local algebra in QFTCS in our setup~\cite{Chandrasekaran:2022cip}. 
According to our quantum equivalence, one can ask how to realize the information loss problem in the view point of IQFT. That may be related to understand the Hawking radiation in the context of IQFT. In this regard, we guess that the CGHS model~\cite{Callan:1992rs} becomes a good test ground for this physics.

\vskip 1cm
\section*{Acknowledgments}
We appreciate conversations and discussions with Dongsu Bak, Seungjoon Hyun, Chanju Kim, Kyung Kiu Kim,  Wontae Kim, Yoonbai Kim, Miok Park, and Driba D. Tolla. 
This work was supported by the National Research Foundation of Korea(NRF) grant with grant number NRF-2022R1F1A1073053(O.K.), NRF-2020R1A2C1014371(O.K. and J.H.), \\ NRF-2020R1C1C1012330(S.-A.P.), 
 NRF-2021R1A2C1003644(S.-H.Y.) and supported by Basic Science Research Program through the NRF funded by the Ministry of Education 2020R1A6A1A03047877(S.-H.Y. and J.H.).


\newpage 

\begin{center} {\Large \bf Appendix}
\end{center}

\begin{appendix}
\section{Isometries in $(1+1)$-dimensional Background}\label{AppA}

In this Appendix, we summarize some formulae used in the main text. For the two-dimensional metric,
\begin{align}
        ds^2=e^{2\omega(x)}(-dt^2+dx^2)\,,
\end{align}
the non-vanishing Christoffel symbols are given by
\begin{equation} \label{}
\Gamma^{t}_{tx}= \Gamma^{t}_{xt}= \Gamma^{x}_{xx} = \Gamma^{x}_{tt}=  \omega'(x)\,,
\end{equation}
where ${}'$ denotes the differentiation with respect to $x$.  
In this geometry, the Ricci tensor and the curvature scalar are given by
\begin{align}
        R_{\mu\nu} = -g_{\mu\nu}~ \omega''\,, \qquad R=-2e^{-2\omega}\omega''\,.
\end{align}

Now we show that the spacetime described by the above metric admits only a time-like Killing vector, excepting dS, AdS, and Minkowski spacetimes which have three Killing vectors.
Killing condition on this background   $\nabla_{(\mu} \xi_{\nu)}=0$ becomes 
\begin{equation} \label{}
\xi^{x} = C(t) e^{-\omega(x)}\,, \qquad  (\xi^{x})^{\Large\boldsymbol{\cdot}}-(\xi^{t})' =0\,, \qquad  (\xi^{t})^{\Large\boldsymbol{\cdot}} + \omega' \xi^{x}= (\xi^{t})^{\Large\boldsymbol{\cdot}}  - (\xi^{x})'  =0\,,
\end{equation}
where $\Large\boldsymbol{\cdot}$ denotes the differentiation with respect to $t$.  This condition leads to 
\begin{equation} \label{}
\frac{\ddot{C}}{C} = \frac{(e^{-2\omega})''}{e^{-2\omega}} = A_{0} = const.  \quad \text{if} \quad C\neq 0\,. 
\end{equation}
When $C=0$, one obtains $\xi^{\mu} = (1,0)$ or $\xi = \partial_{t}$,  up to normalization. In the case of $C\neq 0$, 
we can solve the differential equation  $ \frac{(e^{-2\omega})''}{e^{-2\omega}} = - A_{0}$, which leads to  
\begin{equation} \label{}
e^{-\omega} = %
\left\{ \begin{array}{lll }    
D_{c}\cosh Bx + D_{s}\sinh Bx \,,  & \text{when}   &  A_{0} = B^{2}  >  0    \\
D\cos B(x-x_{0})\,,  &   \text{when}  & A_{0}  = - B^{2} <   0 \\
D_{1}x + D_{2}\,,  &   \text{when}  & A_{0}  = 0
\end{array}  \right.   \,,
\end{equation}
where $D_{c/s}, D, D_{1/2}$, and $x_{0}$ are integration constants. One can check that upper two cases ($A_{0}$ is positive or negative)  correspond to dS$_{2}$ and AdS$_{2}$, respectively. While the last one with $D_{1}=0$ corresponds to the Minkowski spacetime. If $D_{1}\neq 0$, then there is a singularity.  
This computation tells us that there is no other Killing vector except for $\xi = \partial_{t}$ ($C=0$ case) for a generic non-singular metric. 

As an example, let us consider the case of $A_{0} =B^{2}$ with $D_{c}=0$. The independent Killing vectors up to normalization  are obtained as
\begin{equation} \label{}
\xi  = \xi^{t}\partial_{t} + \xi^{x}\partial_{\xi} = %
\left\{ \begin{array}{l}    
\cosh Bx\sinh Bt~ \partial_{t} + \sinh Bx \cosh Bt~ \partial_{x}     \\
\cosh Bx\cosh Bt~ \partial_{t} + \sinh Bx \sinh Bt~ \partial_{x}      \\
\partial_{t}  
\end{array}  \right.    \,.
\end{equation}
One can check that these three Killing vectors form a $SO(2,1)$ algebra.  Indeed, performing the coordinate transformation $r=r_{H}\coth Bx$, we obtain the Rindler wedge of   AdS$_{2}$ geometry.
In the case of $A_{0} =B^{2}$ with $D_{s}=0$, one obtains
\begin{equation} \label{}
\xi  = \xi^{t}\partial_{t} + \xi^{x}\partial_{\xi} = %
\left\{ \begin{array}{l}    
\sinh Bx\sinh Bt~ \partial_{t} + \cosh Bx \cosh Bt~ \partial_{x}     \\
\sinh Bx\cosh Bt~ \partial_{t} + \cosh Bx \sinh Bt~ \partial_{x}      \\
\partial_{t}  
\end{array}  \right.    \,,
\end{equation}
which corresponds to the Killing vectors in the static path of dS$_{2}$ spacetime. 
\vskip1cm

\section{Canonical Quantization in Curved Spacetime} \label{AppB}

In this Appendix, we review the quantization procedure in a $(1+1)$-dimensional curved background~\cite{Birrell:1982ix,Jacobson:2003vx}. We consider the quadratic action \eqref{GraLag2} for the real scalar field with the coupling $\xi$ on a background geometry. The equation of motion of the model \eqref{GraLag2} is given by 
\begin{align}\label{EOM1}
\left(-\Box + m_0^2 + \xi \mathcal{R} \right) \phi = 0, \qquad \Box \equiv \frac{1}{\sqrt{-g}} \partial_{\mu}\left(\sqrt{-g} g^{\mu\nu} \partial_{\nu}\right).
\end{align}
The canonical momentum for the field $\phi$ at a constant time $t$ is read as 
\begin{align}
\pi (t,x) \equiv \frac{\delta S_{{\rm FTCS}}}{\delta \partial_{t} \phi (t,x)} = \sqrt{h} \, n^\mu \partial_{\mu} \phi(t,x), 
\end{align}
where $n^\mu = -\frac{g^{\mu 0}}{\sqrt{-g^{00}}}$ is the unit normal vector to the hypersurface $\Sigma$, and $h$ is the determinant of the induced spatial metric on the surface. In order to quantize the field $\phi (t,x)$, one has to promote the fields, $\phi (t,x)$ and $\pi (t,x)$, to Hermitian operators and require the commutation relation at a fixed time $t$, 
\begin{align}\label{ETCR1}
&[ \phi(t,x), \pi (t, y) ] = i\delta (x-y)\,,
\nonumber\\
 &[ \phi(t,x), \phi (t, y) ] = [ \pi(t,x), \pi (t, y) ]=0.  
\end{align}
The commutation relations in \eqref{ETCR1} are the same forms of those in the Minkowski spacetime. Following the quantization procedure in the Minkowski spacetime, we define an inner product  
\begin{align}\label{brckt1}
\langle \phi_1,\, \phi_2 \rangle = \int_\Sigma d \Sigma^\mu J_\mu,
\end{align}
where $d \Sigma^\mu \equiv n^\mu d \Sigma$ and  $J_\mu \equiv  i \left(\phi_1^* \partial_{\mu} \phi_2 - \partial_{\mu} \phi_1^* \phi_2  \right)$ is a current satisfying the on-shell relation $\nabla^\mu J_\mu = 0$. This bracket is called the Klein-Gordon inner product, and it does not depend on the choice of the spacelike hypersurface $\Sigma$ in the case that the fields decay sufficiently fast at spatial infinity. That is, if we consider another hypersurface $\Sigma'$ at a different time $t'$, we have the relation \begin{align}
\int_\Sigma d \Sigma^\mu J_\mu - \int_{\Sigma'} d \Sigma^\mu J_\mu = \int_{M} d^2 x \sqrt{-g} \,  \nabla^\mu J_\mu = 0,
\end{align}
where $M$ is the manifold bounded by the hypersurfaces, $\Sigma$ and $\Sigma'$. This independence of the hypersurfaces realizes the time-independence of the inner product in the Minkowski spacetime. For complex functions $f$ and $g$ satisfying the equation of motion \eqref{EOM1}, the inner product satisfies the following relations, 
\begin{align}\label{brckt2}
\langle f,\, g\rangle^* = - \langle f^*,\, g^* \rangle = \langle g,\, f\rangle,
\end{align}
which implies $\langle f,\, f^*\rangle =0$. 
In analogy with the quantization of $\phi$ in Minkowski spacetime, we define the annihilation operator related to the function $f$ in terms of the inner product, 
\begin{align}
a(f) = \langle f,\, \phi \rangle,
\end{align}
which is independent of the hypersurface $\Sigma$. Using the properties of the inner product in \eqref{brckt2} and the Hermiticity of the field operator $\phi$, we obtain the Hermitian conjugate of $a(f)$, $a^\dagger(f) = - a (f^*)$. Using these relations and the commutation relations in \eqref{ETCR1}, we obtain the following commutation relations
\begin{align}\label{numop1}
[a(f), a^\dagger (g)] = \langle f,\, g\rangle, \quad [a(f), a (g)] = -\langle f,\, g^*\rangle, \quad  [a^\dagger (f), a^\dagger (g)] =- \langle f^*,\, g\rangle.
\end{align} 
When the complex solution $f$ satisfies $\langle f,\, f\rangle = 1$, by setting $g = f$ in \eqref{numop1} one can easily see that the relations in \eqref{numop1} are nothing but commutation relations of number operators in harmonic oscillator. 
Therefore in order to quantize the scalar field $\phi$ in a curved background we have to find a complete orthonormal basis of solutions to \eqref{EOM1} satisfying the inner product relations
\begin{align}
\langle u_i,\, u_j\rangle = \delta_{ij}, \quad \langle u^*_i,\, u_j\rangle = 0, \quad \langle u_i^*,\, u_j^*\rangle = - \delta_{ij},
\end{align}
where corresponding annihilation and creation operators are denoted by $a_i$ and $a_i^\dagger$, respectively. 
Then one can expand the scalar field $\phi$ as 
\begin{align}
\phi(t,x) =\sum_{i} \left( a_i u_i + a_i^\dagger u_i^*\right)
\end{align}
with the commutation relations $[a_i,\, a_j^\dagger ] = \delta_{ij}$. The Fock space can be constructed by these operators.

\end{appendix}
\newpage 

\center{
Data Availability Statement: No Data associated in the manuscript.
}


\begin{thebibliography}{99}
\bibitem{wtkim}
We would like to thank Wontae Kim for the private discussion about this issue.
\end{thebibliography}
\end{document}